\documentclass{article}

\usepackage{titling}
\usepackage{microtype}
\usepackage{graphicx}
\usepackage{booktabs} % for professional tables
\usepackage{hyperref}
\usepackage{multirow}
\usepackage{xurl} % this solves the url issue in the reference

\usepackage{enumitem}
\usepackage{todonotes}
\usepackage{xspace}
\usepackage{subfig}
\usepackage{array}
\usepackage[normalem]{ulem}
\usepackage{pifont}
\usepackage{multirow}
\usepackage[linesnumbered,ruled,vlined]{algorithm2e}
\usepackage[compact]{titlesec}
\definecolorseries{test}{rgb}{grad}[rgb]{.95,.55,.55}{11,11,17}
\resetcolorseries[10]{test}
\newcommand{\addtodoeditor}[1]{%
    \colorlet{#1}{test!!+!50}
    \expandafter\newcommand\csname#1\endcsname [1]{%
        \todo[color=#1,size=\tiny]{\sffamily\textbf{\uppercase{#1}:}
    ##1}\xspace%
    }
    \expandafter\newcommand\csname#1i\endcsname [1]{%
        \todo[inline, color=#1, size=\tiny]{\sffamily\textbf{\uppercase{#1}:} ##1}\xspace%
    }
}
\addtodoeditor{dr}
\addtodoeditor{bz}
\addtodoeditor{bs}

\definecolor{skyblue}{RGB}{135,206,235}
\newcommand{\helper}[1]{\textcolor{lightgray}{#1}}
\newcommand{\RightTriangle}{\tikz\draw[fill=skyblue!50] (0,0) -- (0.2cm,0.1cm) -- (0,0.2cm) -- cycle;}
\newcommand{\Circle}{\tikz\draw[fill=skyblue!50] (0,0) circle (0.1cm);}
\newcommand{\Square}{\tikz\draw[fill=skyblue!50] (0,0) rectangle (0.2cm,0.2cm);}
\newcommand{\Hexagon}{\tikz\draw[fill=skyblue!50] 
  (0:0.12cm) -- (60:0.12cm) -- (120:0.12cm) -- 
  (180:0.12cm) -- (240:0.12cm) -- (300:0.12cm) -- cycle;}

\newcommand{\RHexagon}{\tikz\draw[fill=skyblue!50, rotate=90] 
  (0:0.12cm) -- (60:0.12cm) -- (120:0.12cm) -- 
  (180:0.12cm) -- (240:0.12cm) -- (300:0.12cm) -- cycle;}

\newcommand{\Octagon}{\tikz \draw[fill=skyblue!50,rotate=22.5] (0:0.12cm) -- (45:0.12cm) -- (90:0.12cm) -- 
  (135:0.12cm) -- (180:0.12cm) -- (225:0.12cm) -- (270:0.12cm) -- (315:0.12cm) -- cycle;}

\newcommand{\cmark}{\ding{51}} % tick
\newcommand{\xmark}{\ding{55}} % cross

\newcommand{\sys}{\emph{ProfInfer}\xspace}

\usepackage[accepted]{mlsys2025}

\mlsystitlerunning{ProfInfer: An eBPF-based Fine-Grained LLM Inference Profiler}

\begin{document}

\twocolumn[
\mlsystitle{ProfInfer: An eBPF-based Fine-Grained LLM Inference Profiler}

% It is OKAY to include author information, even for blind
% submissions: the style file will automatically remove it for you
% unless you've provided the [accepted] option to the mlsys2025
% package.

% List of affiliations: The first argument should be a (short)
% identifier you will use later to specify author affiliations
% Academic affiliations should list Department, University, City, Region, Country
% Industry affiliations should list Company, City, Region, Country

% You can specify symbols, otherwise they are numbered in order.
% Ideally, you should not use this facility. Affiliations will be numbered
% in order of appearance and this is the preferred way.
% \mlsyssetsymbol{equal}{*}

\begin{mlsysauthorlist}
\mlsysauthor{Bohua Zou}{drc,tum}
\mlsysauthor{Debayan Roy}{drc}
\mlsysauthor{Dhimankumar Yogesh Airao}{drc}
\mlsysauthor{Weihao Xu}{tum}
\mlsysauthor{Binqi Sun}{tum}
\mlsysauthor{Yutao Liu}{drc}
\mlsysauthor{Haibo Chen}{csi,sjtu}
% \mlsysauthor{Bauiu C.~Yyyy}{equal,to,goo}
% \mlsysauthor{Cieua Vvvvv}{goo}
% \mlsysauthor{Iaesut Saoeu}{ed}
% \mlsysauthor{Fiuea Rrrr}{to}
% \mlsysauthor{Tateu H.~Yasehe}{ed,to,goo}
% \mlsysauthor{Aaoeu Iasoh}{goo}
% \mlsysauthor{Buiui Eueu}{ed}
% \mlsysauthor{Aeuia Zzzz}{ed}
% \mlsysauthor{Bieea C.~Yyyy}{to,goo}
% \mlsysauthor{Teoau Xxxx}{ed}
% \mlsysauthor{Eee Pppp}{ed}
\end{mlsysauthorlist}

% \mlsysaffiliation{to}{Department of Computation, University of Torontoland, Torontoland, Canada}
% \mlsysaffiliation{goo}{Googol ShallowMind, New London, Michigan, USA}
% \mlsysaffiliation{ed}{School of Computation, University of Edenborrow, Edenborrow, United Kingdom}
\mlsysaffiliation{drc}{Huawei Hilbert Research Center (Dresden), Dresden, Germany}
\mlsysaffiliation{tum}{Technical University of Munich, Munich, Germany}
\mlsysaffiliation{csi}{Huawei Central Software Institute}
\mlsysaffiliation{sjtu}{Shanghai Jiao Tong University, Shanghai, China}

\mlsyscorrespondingauthor{Debayan Roy}{debayan.roy6@huawei.com}
% \mlsyscorrespondingauthor{Eee Pppp}{ep@eden.co.uk}

% You may provide any keywords that you
% find helpful for describing your paper; these are used to populate
% the "keywords" metadata in the PDF but will not be shown in the document
\mlsyskeywords{Machine Learning, MLSys}

\vskip 0.3in

\begin{abstract}
As large language models (LLMs) move from research to production, understanding how inference engines behave in real time has become both essential and elusive. Unlike general-purpose engines such as ONNX Runtime, today’s LLM inference systems offer little operator-level visibility, leaving developers blind to where time and resources go. Even basic questions—is this workload memory-bound or compute-bound?—often remain unanswered.
To close this gap, we develop a fine-grained, non-intrusive profiling framework for modern LLM inference engines, exemplified by llama.cpp but applicable to similar runtime architectures. Built on extended Berkeley Packet Filter (eBPF) technology, our system dynamically attaches probes to runtime functions across multiple layers—without modifying or recompiling the source. It transforms collected traces into rich visualizations of operators, graphs, timelines, and hardware counter trends, exposing how dense inference, Mixture-of-Experts routing, and operator offloading behave in practice.
With less than 4\% runtime overhead and high profiling fidelity, our framework makes LLM inference both transparent and diagnosable, turning performance profiling into a practical tool for optimization, scheduling, and resource-aware deployment.
\end{abstract}
]
\printAffiliationsAndNotice{}  % leave blank if no need to mention equal contribution
% \printAffiliationsAndNotice{\mlsysEqualContribution} % otherwise use the standard text.

\section{Introduction}
% \bsi{This is a new version of intro drafted by binqi, feel free to change anything or merge useful parts into the main branch}

Large language models (LLMs) are reshaping how we interact with computers by powering conversational agents, writing assistants, multilingual translation services, travel-planning aids, and personalized tutors. Underpinned by transformer architectures that process tokens while computing inter-token attention relationships, LLMs deliver strong natural-language understanding and generation capabilities. Deploying LLM inference on mobile and edge devices rather than relying exclusively on cloud infrastructure opens important opportunities, including offline operation, lower latency, enhanced user privacy, and reduced network dependence~\cite{pangupi, apple-on-device}.

Enabling on-device LLM inference is far from trivial. Even when employing highly optimized inference engines, the computation and memory demands of LLMs are orders of magnitude greater than those of traditional mobile neural networks. A model may generate text token-by-token, rather than in a single forward pass, and its inference proceeds through distinct phases: a \emph{prefill} phase that is typically compute-bound, and a \emph{decode} phase that is often memory- or bandwidth-bound. Meanwhile, mobile devices impose stringent limits on processor power, memory size, caching behavior, thermal dissipation, and energy budgets. As recent studies highlight~\cite{chen2025characterizing,lipalmbench}, deploying LLMs on mobile hardware requires careful balancing of model size, memory management, and accelerator usage. In such constrained and heterogeneous environments, accurate and low-overhead performance visibility becomes essential for identifying inefficiencies and guiding optimizations.

Profiling tools are crucial for understanding and optimizing ML workloads: by exposing operator-level timings, memory usage, and threading behavior, they enable developers to pinpoint performance bottlenecks. General-purpose ML inference engines such as ONNX Runtime~\cite{onnxruntime_prof} and TensorRT~\cite{tensorrt_profiler} provide built-in profilers for operator execution time, memory allocation, and thread utilization, aiding optimization in latency- and resource-sensitive settings. Yet for LLM inference engines, especially those targeting on-device and edge scenarios, we observe a striking lack of \emph{fine-grained} and \emph{non-intrusive} profiling support. Existing profilers often require recompilation or runtime instrumentation, offer only coarse-grained metrics such as throughput or token rate, and fail to expose critical dimensions of execution such as dynamic operator graphs, per-thread scheduling, hardware-counter events, and phase-specific behaviors. Besides, some hardware-based profilers typically rely on support from the underlying hardware and often introduce non-negligible overhead.

While extended Berkeley Packet Filter (eBPF) has recently been used for tracing deep learning performance~\cite{chu2025einfer,yang2025egpu,craun2024eliminating}, it remains largely decoupled from LLM execution semantics. Existing work rarely correlates low-level hardware metrics with high-level operator semantics, nor has it been adapted to mobile or edge environments where runtime overhead and observability constraints are more stringent. Consequently, developers lack the necessary observability to understand how quantization, KV-cache reuse, accelerator offloading, or memory bandwidth constraints affect end-to-end latency and efficiency on-device. 

To address these limitations, we present \sys, a profiling framework tailored to on-device LLM inference. The design of \sys\ emphasizes three key attributes. First, it offers \emph{fine-grained observability} by capturing every forward pass and operator invocation (e.g., matrix multiplication, attention, softmax) along with tensor dimensions, operator types, and the dynamically constructed computational DAG. Second, it ensures \emph{non-intrusive instrumentation} through the use of eBPF probes that attach to runtime functions without requiring source modification, enabling easy deployment on Linux-based mobile and edge operating systems. Third, it supports \emph{hardware-counter integration and visual analytics}, collecting per-operator performance monitoring counter (PMC) data, such as cache misses, memory accesses, and thread execution, and presenting them through timeline views, DAG visualizations, and per-operator plots. 

\begin{figure}
    \centering
    \includegraphics[width=0.9\columnwidth]{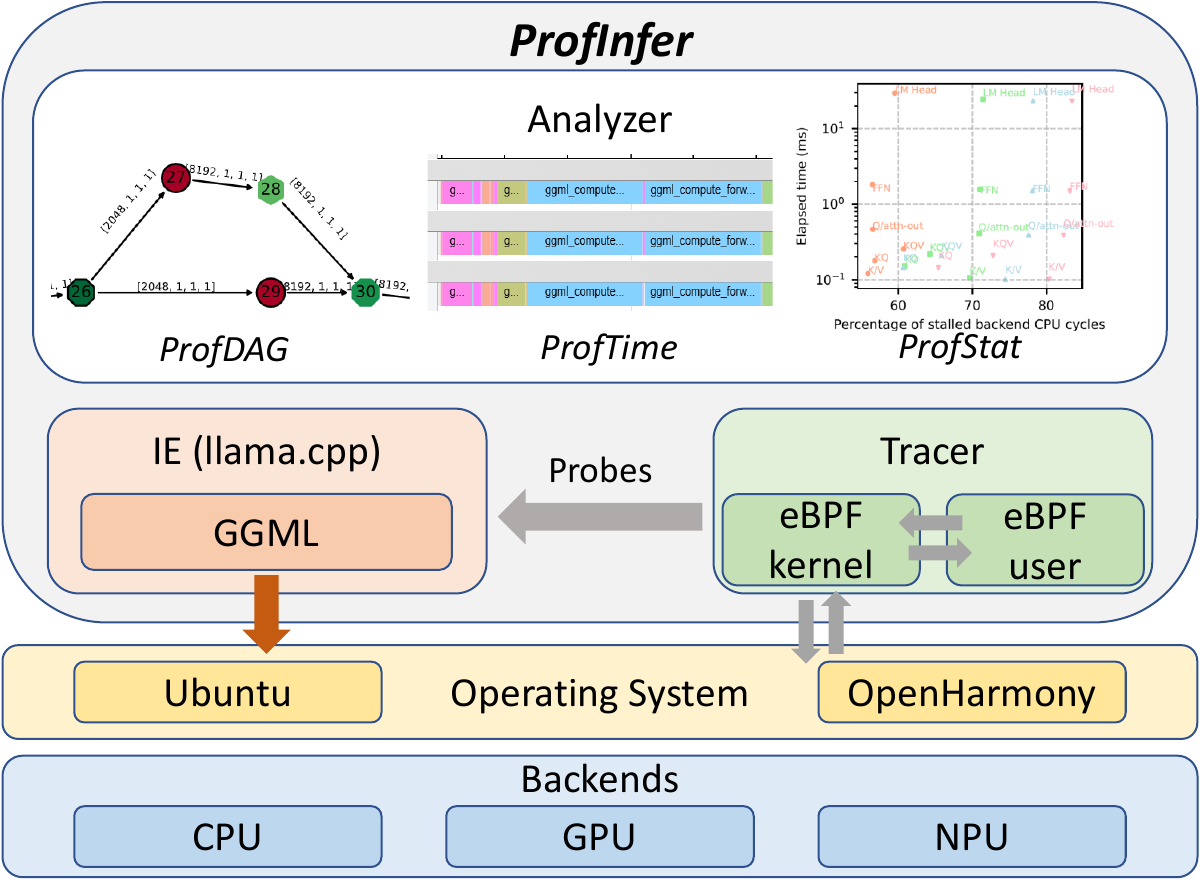}
    \caption{High-level design of {\sys} comprising eBPF-based tracers followed by trace analyzers.}
    \label{fig:overview-profinfer}
\end{figure}

We implement \sys\ over the llama.cpp~\cite{llamacpp} inference engine, with a high-level overview illustrated in Figure~\ref{fig:overview-profinfer}. 
% Towards developing {\sys}, we study the software architecture and implementation of llama.cpp and identify several crucial functions in its different layers. 
% \bs{Maybe tune the below texts to describe the design shown in the figure, instead of listing the details, which seem to be unconnected to each other}
Before starting the workload, the \emph{tracer} attaches several probes to the runtime libraries and enables relevant system tracing events. Online, the \emph{tracer} continuously gathers and logs the submitted buffer from each probe handler in the kernel space and conditionally switches off the unnecessary probes. Offline, the \textit{analyzer} parses the collected results, identifies the computation and the backend types, and then provides three different representations of the structural and performance metrics at the token, graph, and operator level. 

% For example, (i)~\emph{llama\_decode} can be parameterized to run a forward pass through the model, whereas (ii)~\emph{ggml\_compute\_forward} executes each operator in the computational graph. 
% Further, we leverage eBPF to attach \emph{u(ret)probes} to the identified functions. 
% We note that eBPF is a Linux kernel technology that enables running sandboxed programs in kernel space without the need to modify or recompile the kernel source. 
% The attached probes can report the timings of function calls and returns. 
% Besides, in the probe handlers, we parse the function call arguments and return values to extract valuable information, e.g., the operator type and more details about input and output tensors.

The main contributions of this paper are as follows:
\begin{itemize}
    \item We propose \sys, an eBPF-based profiling framework for LLM inference that is fine-grained, non-intrusive, and lightweight, making it suitable for both mobile and edge environments.
    \item \sys enables comprehensive visibility across the entire inference pipeline—covering each forward pass, computation graph, operator execution, and processor thread, while collecting a rich set of performance metrics through hardware counters.
    \item It offers intuitive performance analytics, including timeline views, DAG visualizations, and op-level plots that correlate model structure with hardware behavior.
    \item We demonstrate how \sys facilitates analysis of compute and memory bottlenecks, KV-cache effects, workload interference, backend performance divergence, and the dynamic characteristics of MoE models.
    % and the trade-offs in batched and parallel inference on mobile devices.\bs{check if anything is missing or needs updated}
\end{itemize}

% \bs{If there is space, add a short paragraph to introduce the organization of the paper}

% The remainder of the paper is organized as follows. 
% Section~\ref{sec:background} introduces necessary preliminaries. 
% Sections~\ref{sec:tracer} and~\ref{sec:analyzer} present the tracers and analyzers developed in \sys, respectively. 
% Section~\ref{sec:eval} shows experimental results with valuable insights. 
% Section~\ref{sec:discuss} discusses the applicability of \sys, and Section~\ref{sec:related} reviews relevant existing works. 
% Section~\ref{sec:conclusion} concludes the paper with future directions. 
% \bs{add back the organization para if there's enough space}

% \input{sections/intro}
% \newpage

\section{Background}
\label{sec:background}
%4 columns

\subsection{Large language model (LLM)}
Large Language Models (LLMs) are mostly transformer-based models leveraging the self-attention mechanism to capture contextual information~\cite{vaswani2023attentionneed}. After GPT~\cite{radford2018improving} is released, the decoder-only LLM model is primarily adopted for generative tasks and exhibits superior general intelligence. Recently, an increasing number of LLMs have been open-sourced, and their performance now rivals that of many closed-sourced LLMs.

Diving into the architecture of a decoder-only LLM of LLaMA family~\cite{grattafiori2024llama}, it inherits the basic design of GPT, stacking multiple transformer layers of the identical shape. Each layer is composed of a self-attention layer and a feed-forward network (FFN). In particular, LLaMA differs by adopting grouped query attention (GQA)~\cite{ainslie-etal-2023-gqa}, rotary positional embedding, root-mean-squared layer-normalization (RMS Norm), and SwiGLU as the activation function. Other families of LLMs, such as Qwen~\cite{bai2023qwen}, Phi, Gemma, also have a unique architectural design. Except for the training dataset and the maximum context length, the most fundamental difference lies in the structural dimension determined by several hyperparameters, such as the number of layers, hidden size, and number of attention heads. 

% These hyperparameters, like the number of layers, hidden size, and number of attention heads, can further lead to different computation and memory access patterns during the inference process.
\subsection{LLM inference and inference engine}
\subsubsection{LLM inference}
Most decoder-only LLMs employ an auto-regressive manner to do the inference. After processing the prompt in the prefill stage, the LLM generates only one token at each decoding iteration to ensure causality, until it reaches a pre-set maximum length or a special token like \texttt{<eos>}. 
The performance of an LLM is commonly evaluated using two key metrics: the time to the first token (TTFT) in the prefill stage, and the decoding speed or period, i.e., tokens per second (TPS) or time per output token (TPOT), in the decoding iteration. Both two metrics are determined by the model size, architecture, and the hardware configuration. 
% At a finer granularity, the types, scales, and the topology of the operators play a more critical role in determining the overall performance.

By exploring the sparsity inside LLMs, more architecture-wise innovations, such as dynamic pruning (PowerInfer~\cite{song2024powerinfer}), and mixture-of-experts (MoE)~\cite{MoE}, can further accelerate on-device LLM inference by activating only a part of the weights online. Besides, speculative decoding can boost the overall decoding speed by running a smaller draft LLM while using a bigger target LLM for verification after a certain number of tokens have been generated~\cite{leviathan2023fast}. The uncertainty introduced by these dynamic workloads can also lead to performance fluctuations. For example, in MoE models, the frequency with which the same experts are activated affects runtime memory and disk I/O overhead, which also impacts the inference performance.

% , even if the entire model size is bigger than the cache of DRAM. However, due to the stochastic online activated neurons or layers, the cache locality cannot be fully utilized, and model weights need to be frequently loaded from the disk. Predictive dynamic pruning can pre-fetch the data, but the accuracy is constrained with the complexity of the activation functions.
% \begin{itemize}
%     \item explain the architecture of LLMs
%     \item explain how LLM inferences work at a high level including the different nature of prefill and decode
%     \item explain the importance of inference timings, i.e., prefill and decode speeds
% \end{itemize}

\subsubsection{LLM inference engine}
Unlike conventional neural networks that only perform a forward pass, LLM inference also handles tokenization, sampling, and KV cache management. As LLM inference usually requires much more computing and memory resources, some light-weight designs are usually adopted in the LLM IEs. vLLM~\cite{kwon2023efficient} utilizes the paged attention to optimize the memory allocation for the KV cache across different batches, while it is not suitable for mobile scenarios where the batch size is usually 1. MNN-LLM~\cite{wang2024mnn} and mlc-llm~\cite{mlc-llm} are two representative IEs for mobile devices that support different types of models and hardware. The former is a C++-based library that optimizes the memory layout and access pattern and also utilizes the combined quantization. The latter is based on Apache TVM~\cite{chen2018tvm} and implements operator- and graph-level optimization during compilation.

{\sys} mainly focuses on llama.cpp~\cite{llamacpp}, one of the most widely used open-source IEs, which has a bunch of well-known front-ends, like Ollama~\cite{ollama2023}, many bindings in other programming languages, and also many derivatives, such as PowerInfer~\cite{song2024powerinfer}. As a C/C++ based IE, llama.cpp provides versatile functionalities, such as \textit{llama-server}, speculative decoding, and multimodal model inference. Moreover, llama.cpp is built upon GGML, a machine learning runtime library that executes the graph and operator computation, which supports deployment across heterogeneous backends, including CPU, GPU, and certain types of NPU.
% \begin{itemize}
%     \item list a few engines that can serve at edge devices. (could refer to~\cite{park2025survey})
%     \item emphasize that llama.cpp is the infrastructure of many front-ends (UIs). Also, it has so many bindings in other programming languages.
%     \item explain the architecture of llama.cpp
% \end{itemize}

% \subsubsection{Supported model architectures and inference technologies}
% \begin{itemize}
%     \item supported models (dense models, MoE models, VLMs, quantization)
%     \item explain different technology, e.g., batching, speculative decoding, MoE, that we will support profiling for; say that llama.cpp supports such technology and later we should point out the specific thing in the framework tailored for them.
% \end{itemize}
% % Inference technologies (optimization methods) that edge devices can benefit. But for the others (e.g. numa-aware, flash attention), we could point out that these cannot help accelerate inference.
% \begin{itemize}
%     \item KV cache: trade-off between computation and memory access.
%     \item CPU: multi-threading, SIMD, BLAS, already outperform accelerators due to the memory bandwidth bottleneck.
%     \item Multi-backend (heterogeneous computing)
% \end{itemize}

\subsection{Extended Berkeley Packet Filter (eBPF)}
\begin{table}[]
    \centering
    \resizebox{\columnwidth}{!}{
        \begin{tabular}{c|cc}
            \toprule
              & BPF APIs &  Description \\
             \midrule
             \multirow{8}{*}{\rotatebox{90}{Kernel space}} & \small\texttt{bpf\_ktime\_get\_ns} & Get the timestamp \\
             & \small\texttt{bpf\_get\_current\_pid\_tgid} & Get pid and tid  \\
             & \small\texttt{bpf\_get\_smp\_processor\_id} & Get cpu id \\
             & \small\texttt{bpf\_probe\_read\_user(kernel)} & Read user/kernel address space \\
             % \texttt{bpf\_probe\_read\_kernel()} & Read kernel address space \\
             & \small\texttt{map.lookup} & Query a BPF map \\
             & \small\texttt{map.update} & Update a BPF map \\
             & \small\texttt{map.perf\_read} & Read the value of a PMC \\
             % \texttt{ringbuf\_reserve()} & Reserve a ring buffer \\
             % \texttt{ringbuf\_discard()} & Discard a ring buffer \\
             & \small\texttt{ringbuf(perf)\_submit} & Submit a ring(perf) buffer \\
             \hline
             \multirow{5}{*}{\rotatebox{90}{User space}} & \multirow{2}{*}{\small\texttt{attach\_u(ret)probe}} & Attach probes when a user-space \\ & & function is invoked or returned \\
             & \small\texttt{attach\_tracepoint} & Attach a kernel trace point\\
             & \small\texttt{open\_perf\_event} & Create a fd to monitor PMC \\
             & \small\texttt{perf(ring)\_buffer\_poll} & Poll the submitted perf(ring) buffer\\
             \bottomrule

        \end{tabular}
    }
    \caption{A list of \texttt{BCC}-offered APIs used in our \texttt{BPF C} and \texttt{Python} programs. \texttt{libbpf} also offers equivalent ones.}
    \label{tab:bpf_apis}
\end{table}
\emph{eBPF} is a powerful technology that enables to run sandboxed programs in the kernel space, thereby extending kernel functionality without requiring to modify and recompile the kernel source code~\cite{rice2023learning}.
It was first supported by Linux---introduced in 2014---but later expanded to macOS and Windows. 
Since it offers avenues to hook into kernel events and system calls, and to attach probes to user-space function calls and returns, it can be used to develop sophisticated system profiling tools~\cite{cassagnes2020rise}.

In this work, we use two popular frameworks, \texttt{libbpf} and \texttt{BCC} (BPF Compiler Collection), to implement eBPF programs.  
\texttt{BCC} provides Python bindings and is ideal for rapid prototyping, hence, we use it to implement {\sys} over Ubuntu. 
However, since OpenHarmony does not support a Python environment by default, we implement {\sys} over it using a C library \texttt{libbpf}.
In Table~\ref{tab:bpf_apis}, we summarize the BPF APIs used in our \texttt{BCC}-based implementation of \sys's tracer, as detailed in Section~\ref{sec:tracer}. 
% We encourage to refer to them while reading different implemented features of {\sys}'s tracer in Section~\ref{sec:tracer}.

%\begin{itemize}
%    \item besides the general intro about eBPF, we should talk about the APIs that are useful, e.g., attaching probes, reading arguments, and performance counter monitoring
%    \item we should talk briefly about bcc and libbpf (only if we use it for OpenHarmony)
%\end{itemize}

% \newpage 
\section{LLM Tracer in ProfInfer}
\label{sec:tracer}
% 1-2 columns
% ProfInfer includes both tracing and analytics

\begin{figure}
    \centering
    \includegraphics[width=\columnwidth]{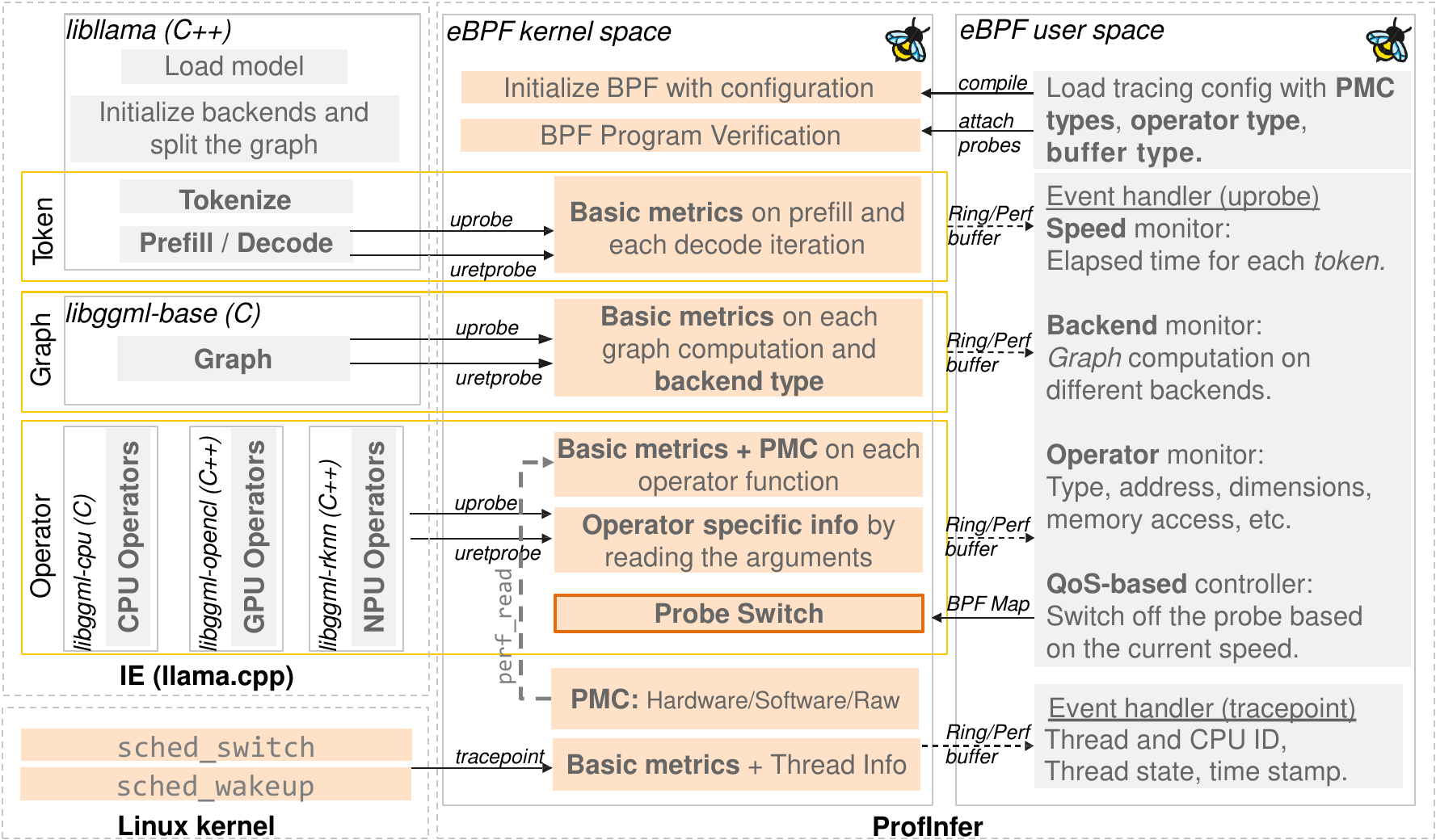}
    \caption{An operational overview of {\sys}'s tracer for LLM inferences over llama.cpp. Basic metrics refer to type of the probe, time stamp, thread ID and CPU ID.}
    % \bzi{I have modified this figure a bit. The text should be adjusted a bit as well.}
    \label{fig:overview-tracer}
\end{figure}

Figure~\ref{fig:overview-tracer} gives an overview of {\sys}'s tracer. 
A part of the tracer operates in user space while the other part runs in kernel space.
Further, it attaches probes to several functions of llama.cpp to enable tracing LLM inferences at different granularities.
The probe handlers are implemented to collect runtime information about an inference facilitating both architectural and performance analyses. The remainder of this section will explore the implementation and various features of the tracer in greater detail.

\subsection{{\sys} in user space}
{\sys} implements a user-space application that is primarily responsible for compiling, launching, and verifying the BPF program. 
It specifies the compilation flags---according to a user-defined configuration file---for the BPF program, thereby filtering out unnecessary information to reduce tracing overheads. 
For example, a flag can be set to retrieve tensor dimensions when needed for an analysis.
The user-space application further attaches probes (uprobes, uretprobes, and tracepoints) to target functions.
It subsequently polls for the buffers submitted by the probe handlers in the kernel space.
When it receives a buffer, it will asynchronously process the data based on the probe type in the event handler.
It also conditionally writes back into the BPF maps to control the probes and the information to be traced.
That is, when it detects a degradation in the inference performance, i.e., the overall decoding speed is under the Quality of Service (QoS) requirement, it can disable certain tracing features to reduce the overhead and improve the speed.

%
%In the eBPF user space, {\sys} is primarily responsible for compiling, launching, and verifying the BPF program. 
%When monitoring, the user-space part should handle the buffer submitted from the kernel space, and conditionally write back into the BPF maps to control the probes and the types of metrics.

%Based on the user configuration, {\sys} filters out unnecessary information based on user configurations by specifying corresponding flags during compilation---for example, whether to retrieve tensor addresses and dimensions---in order to reduce the overhead. 
%{\sys} also attaches probes (uprobes, uretprobes, and tracepoints) to target functions and subsequently polls for the buffers submitted from the kernel-space {\sys}. When it receives a buffer, {\sys} will asynchronously process the data based on the probe types in the event handler. When it detects the degradation of the performance, i.e., the overall decoding speed is under the QoS requirement, {\sys} will write back to the corresponding BPF maps to disable the tracing on some types of operators.

\subsection{{\sys} in kernel space}
{\sys} implements several probe handlers that run in the kernel space.
Each of them traces the PID of the thread executing the probed function, the CPU where the thread is running, and the timestamp when the handler is invoked.
Further, some of the handlers examine the arguments of the probed function, and by parsing them, they can infer the state of the inference.
These arguments are typically C structs or pointers to C structs containing information about the current operator including its type, input and output tensors, among others.
To log the traced data, a handler submits a \emph{ring buffer} or a \emph{perf buffer} to {\sys}'s user-space application where the former can enhance efficiency but at the cost of unreported missing events.

%In the eBPF kernel space, {\sys} defines several probe handlers on the targeting functions. In addition to reading some basic metrics with the BPF helper functions, such as thread ID, CPU ID, and timestamp, these handlers can also read the arguments of specific functions. By parsing these arguments, which are typically C structs or pointers to C structs, {\sys} can get more information about the graphs and operators, e.g., the operator type and the tensor dimension. To log the parsed results, {\sys} submits a ring buffer or a perf buffer to the user-space part, the former of which could enhance efficiency, but can potentially cause some missing events without reporting their counts to the user.

\subsection{Multi-granularity eBPF tracing}

\begin{table}[]
    \centering
    \resizebox{\columnwidth}{!}{
    \begin{tabular}{p{4cm}|c|c|c}
    \toprule
    Probed functions & Type & Level & Traced information \\
    \midrule
    \small{\texttt{llama\_decode}} & u(ret)probe & Token & batch size \\
    \small{\texttt{ggml\_backend\_graph \_compute\_async}} & u(ret)probe & Graph & backend type\\
    \small{\texttt{ggml\_compute\_forward}} & u(ret)probe & OP(CPU) & tensor info, PMC\\
    \small{\texttt{ggml\_cl\_compute\_forward}} & u(ret)probe & OP(GPU) & tensor info\\
    \small{\texttt{ggml\_rk\_compute\_forward}} & u(ret)probe & OP(NPU) & tensor info\\
    \midrule
    \texttt{sched\_switch} & tracepoint & kernel & thread IDs\\
    \texttt{sched\_wakeup} & tracepoint & kernel & thread IDs\\
    \bottomrule
    \end{tabular}
    }
    \caption{Probes in {\sys}. The timestamp, the thread ID, and the CPU ID are logged in all probe handlers.}
    \label{tab:probes}
\end{table}

Considering that llama.cpp is an open-source IE, we traverse its logic to identify specific functions to probe based on our requirements of examining its performance at different granularities, i.e., with respect to each token, computational graph, and operator. Nevertheless, we note that it is a vast and complex project, which makes this non-trivial. Besides the IE, we trace a few kernel events that are relevant to thread scheduling.    
Table~\ref{tab:probes} summarizes the important details of the probes we have designed and implemented.

%Given llama.cpp as an open-sourced IE, identifying the user-space functions that are doing token generation, as well as graph and operator computing, is critical, feasible, but not trivial, to empower the profiling tool. Besides, we also pick several syscalls that are relevant to thread behavior and mmap. Table~\ref{tab:probes} summarizes the information of all the probes of this paper.

\subsubsection{Token-level tracing}
Profiling an inference running over llama.cpp at the token level covers the prefill stage and each decoding iteration.
In libllama, \texttt{llama\_decode} is the generic function that is invoked to generate the next token both during prefill and decode stages. Embeddings of the tokenized input are passed to it. 
%In libllama, \texttt{llama\_decode} is the generic function for both stages, by giving one or a batch of the embeddings of the tokenized input. 
By attaching uprobe and uretprobe to this function, we obtain the timestamps of its invocation and the corresponding return; the difference between the timestamps gives TTFT during prefill and TPOT during decoding.
In this work, we use these values during runtime to dynamically adjust the tracing overheads based on the decoding speed QoS requirement (e.g. 5 tokens/s) by switching the probes on or off.
Besides, a future work is to use them to adjust resource allocations to inferences based on timing requirements especially when there are parallel workloads.
% \textcolor{red}{Further, we read a specific argument of \texttt{llama\_decode} to get the \emph{batch size}. That is, TTFT and TPOT can be analyzed with respect to batch size, which is especially useful in \emph{continuous batching} scenarios \cite{}.}
%cite a work on continuous batching
% \bz{Parsing batch size is not done yet, but it is doable to parse the second argument \texttt{llama\_batch} of it.} 

%Besides, the user-space part relies on these metrics to control the enabling and disabling of probes and metrics in the kernel space.

\subsubsection{Graph-level tracing}
To manage heterogeneous backends, llama.cpp adopts graph partitioning and a backend scheduling strategy within the GGML library.
That is, if multiple backends are selected to run an LLM inference, several computational graphs are constructed during the initialization phase such that all operators in a graph run using the same backend.
For example, if a certain number of transformer layers are to be offloaded to a GPU (e.g., using a CUDA backend) due to its restricted memory capacity, two or more graphs are created depending on the selected layers for offloading.
%Supposing only the CPU is chosen as the backend, only one sub-graph could be generated, and graph-level tracing should reveal a similar behavior to what token-level tracing does. 
%For the other backends, llama.cpp partitions the graph in the initialization phase, based on the number of offloaded layers specified by the user, as well as whether the operators are supported by the backend. 
To compute the execution time of each graph in a forward pass, we attach uprobe and uretprobe to the function \texttt{ggml\_backend\_graph\_compute\_async} and collect timestamps from their handlers.
Such information can be useful to discern as well as optimize the graph partitioning and scheduling heuristic in llama.cpp.
Further, the uprobe handler extracts the value of \texttt{guid}, a 16-digit backend identifier, from a function argument.

%{\sys} could get the value of \texttt{guid} from the argument by using \texttt{bpf\_probe\_read}, a 16-digit number, which is the unique identifier of the backend. Thus, tracing the upper-level graph computation functions enables observation of the graph partitioning strategy and the online scheduling behavior across multiple backends.

\subsubsection{Operator-level tracing}\label{subsubsec:op-trace}
For a given backend, we identify the function that executes the logic to identify the type of an operator and, correspondingly, call its low-level implementation.
Currently, we support CPU and OpenCL backends, along with a self-developed Rockchip NPU backend, while extending to other backends remains future work.
The identified function in each of our target backends is listed in Table~\ref{tab:probes}.
We attach probes to the function to collect timestamps of its invocation and return, thereby obtaining the execution time of each tensor operation.
We also parse \texttt{ggml\_tensor}, the second function argument and a C struct.
By traversing this data structure using eBPF APIs, we can extract information about the name and type of an operation, e.g., \texttt{ffn\_out-0} and \texttt{GGML\_OP\_MUL\_MAT} (the output matrix multiplication of the feedforward network in the first transformer layer), as well as the pointer address and dimension of source and target tensors.
Such trace data enable to construct the computational graph and operator schedules as well as analyze performance of each operator during an LLM inference.

%Operator-level tracing represents the finest level of tracing that {\sys} can achieve. 
%By tracing all the operators across the entire network, it can capture both the overall DAG topology and execution order, as well as each operator's type, tensor dimensions, and performance characteristics.
%The functions for the operator forward-pass are defined in different backend libraries, which are listed in Table~\ref{tab:probes}, but always take the C struct \texttt{ggml\_tensor} as the second argument, which is a common interface shared among different graphs and operators. The operator type, source tensor address, and output dimension are all defined inside, from which {\sys} can retrieve the entire topological architecture of the LLM.

\subsubsection{Scheduler tracing}
Besides user-space tracing, we hook into two tracepoints in Linux scheduler, i.e., \texttt{sched\_switch} and \texttt{sched\_wakeup}.
The handlers of these tracepoints can read the state of a thread, i.e., running, runnable, or idle.
We apply filters to only trace the events related to the inference threads otherwise the overheads might become unacceptable as there can be numerous such events.
%This tracing feature helps to analyze how the computational load is distributed among the inference threads.
This tracing feature helps to identify interfering parallel workloads as well as to study their impact on an inference.

%to understand the threading behavior of llama.cpp. A key motivation for this is that the SoTA mobile devices usually have multi-core CPUs, even sometimes multi-core NPUS, e.g., 3-core NPU on RK3588. By tracing these syscalls, we could identify when each thread is running, runnable, or idle. Thus, we could analyze how efficient the underlying OS schedules the inference task.

\subsection{Enabling PMCs}
\begin{table}[]
    \centering
    \resizebox{\columnwidth}{!}{
        \begin{tabular}{cccc}
            \toprule
            PMC & Type & Description & Unit \\
            \midrule
            l3d\_cache\_refill & per-core & Read from memory (A76) & 64B\\
            % l3d\_cache\_rd & per-core & Read from memory(A55) & 64B\\
            mem\_access\_wr & per-core & Write into memory(A76) & 16B\\
            \hline
            major-faults & software & Major page faults & 1 Page\\
            \hline
            cycles & hardware & CPU cycles & 1 Cycle\\
            idle-backend-cycles & hardware & stalled CPU cycles at backend & 1 Cycle\\
            \bottomrule
            
        \end{tabular}
    }
    \caption{A summary of the PMCs to reflect memory access, computational efficiency, and disk IO efficiency.}
    \label{tab:pmc}
\end{table}
Considering that PMCs offer low-level insights into CPU behavior with respect to computations and memory accesses, they are essential for profiling and optimizing software performance.
Hence, in {\sys}, we offer the possibility for opening and reading a PMC at the operator level.
In Table~\ref{tab:pmc}, we list a few meaningful PMCs that we can trace.

Let us consider an example where we want to determine the amount of data fetched from DRAM for an operator.
Per-core \texttt{l3d\_cache\_refill} is a useful PMC for that \cite{pradhan2025predictable}.
We use \texttt{open\_perf\_event} in the user-space program to open and configure a file descriptor for each thread and via it we can read~\texttt{l3d\_cache\_refill} counter per thread.
Now, in the handlers of uprobe and uretprobe to \texttt{ggml\_compute\_forward}, we read the counter using \texttt{perf\_read}. The difference between the two values gives how many times L3 data cache was refilled with 64~B of data by an inference thread for the operator. Summing these values over all inference threads can indicate the total amount of data fetched from DRAM for the operator. 
%This calculation is acceptable only under the assumption that the target CPU cores do not run any other tasks besides the inference.
%Nevertheless, when this assumption does not hold, we expand the handler to \texttt{sched\_switch} tracepoint to read the counter every time an inference thread is preempted and continued-after-preemption between \texttt{ggml\_compute\_forward} call and return. 
%That is, we skip counting the cache line fetches when the inference thread is not running.

We note that such a PMC-based analysis can assist in identifying resource under-utilization or boundedness and other performance bottlenecks for each operator.

\comment{
On mobile platforms, memory bandwidth improvements have lagged behind the growth of computing capabilities. Especially, the decoding stage of LLM inference is dominated by the general matrix vector multiplication (GEMV), which is memory-access intensive~\cite{wang2025neuralink}. Normally, there are several pre-defined \texttt{perf} events that could indicate the last-level cache misses, while they might not be accessible or accurate across different platforms based on the complicated design of the memory hierarchy. To accurately profile the memory bandwidth usage at the operator level, {\sys} reads several per-core PMCs for CPU-backend LLM inference, e.g. \texttt{L3D\_CACHE\_REFILL}~\cite{pradhan2025predictable}.

Moreover, some pre-defined software and hardware events are also adopted to characterize other behaviors, e.g., \texttt{stalled-cycles-backend} can reflect the computational efficiency. Besides, \texttt{major-faults} is used to identify the I/O cost at the storage, as llama.cpp can employ \texttt{mmap} well, especially for the large LLMs that cannot be fitted into the DRAM.
By using the \texttt{open\_perf\_event} and \texttt{BPF\_PERF\_ARRAY}, {\sys} can also read them at any probe handler, and append them to the output buffer.
}

\subsection{Tracing expert activations in MoE models}
When running an inference based on an MoE model, different experts can be activated in the FFN of a transformer layer during each decoding iteration.
Expert activations are determined by a gating mechanism, which is usually implemented by a \emph{softmax} or a \emph{sigmoid} operator followed by a \emph{top\_k} operator. 
The result is a tensor of length $k$ comprising the IDs of $k$ activated experts.
Based on this understanding, we have attached a uprobe to the function \texttt{ggml\_compute\_forward\_mul\_mat\_id} where the probe handler reads the $k$ expert IDs by traversing the third source \texttt{ggml\_tensor} using two-level pointer dereferencing.
We note that the DRAM of a mobile device is typically insufficient to accommodate all available experts. 
Hence, when an activated expert is not in the DRAM, it has to be fetched from the storage, which increases operator's execution time.
That is, traced information related to expert activations enables accurate analysis of performance data at the operator level.

%MoE models can activate different experts across different decoding iterations at the FFN of each transformer layer. Thus, when \texttt{mmap} is enabled and some new experts are activated, it is likely that llama.cpp needs to fetch the data from the storage. In particular, the size of the DRAM of a mobile platform is usually insufficient to load an entire MoE model, even if it is quantized. {\sys} also provides the capability of tracing the IDs of activated experts on the run, which can help analyze the relationship between how frequently the same experts are activated and the performance of each decoding iteration.

%Which experts are activated is decided on the run by the "gating" operator, which could be a \textit{softmax} or a \textit{sigmoid}. The selection results are also saved as intermediate tensors with the length of the number of used experts after doing \textit{top\_k} and \textit{argsort}. Then, the IDs of activated experts serve as a third input of the function \texttt{ggml\_compute\_forward\_mul\_mat\_id} to process the FFN. After deriving the layout of the ID information in memory from the source code, {\sys} can parse it through two-level pointer dereferencing. 

\comment{
\begin{itemize}
    \item illustrate the software architecture of our framework
    \item show that it has several probes (and probe handlers) distributed in different layers (parts) of llama.cpp
    \item show that it has conditional message passing mechanism by using perf buffer/ring buffer to 
    \item show that it has trace processors (scripts or functions in one script) for parsing the trace files to obtain operator graph, variation across iterations (prefill and other decode iterations), variations across different operators, MoE specific, accelerator-specific, and so on.  
    \item it outputs data that can be visualized using external popular tools (chrome tracing, Python libraries like graphviz, etc.). 
\end{itemize}
}

% \input{sections/trace}
% \newpage

\section{LLM Trace Analyzers in ProfInfer}
\label{sec:analyzer}
% 3-4 cols
Taking an LLM and the hardware configuration as input, {\sys} generates the raw tracing data and then, through further analysis, produces the following three types of views. 

\subsection{Profiled DAG (ProfDAG) view}
To reduce the memory footprint of the model, llama.cpp compresses the model into \texttt{gguf} format, which excludes any structural information. 
% {\sys} supports online tracing and extracts the topology of the DAG along with the profiling results for each node, i.e., each operator.
Instead, ProfDAG view can illustrate the topology of the workload in a DAG along with the profiling results for each operator, after analyzing the tracing data. 
The concrete process of ProfDAG retrieval is presented in Algorithm~\ref{alg:profdag}. 

% The input $df_{raw}$ is a structured table, each entry of which contains the information of a buffer submitted by uprobe and uretprobe. The input $n_{tar}$ represents which iteration needs to be parsed. It first categorizes the original data into several tables by tid, and then filters out the operator-level tracing events of the targeting iteration in temporal order (lines 1-4).
We first categorize the raw data by different thread IDs, and filter the targeting iteration only with all the operator functions (lines 1-4).
Subsequently, it iterates over each starting event (with a step of 2), and assigns the corresponding attributes and profiling metrics to the node (lines 5-14). 
The elapsed time should be calculated between the first invocation and the last return of the operator functions across multiple threads, while the difference of PMCs should be calculated on each thread and summed up. The type, name, and dimension of the operator inside the $'info'$ help identify each operator, while the execution order reflects the topological sort generated by the llama.cpp. 

\begin{figure}
    \centering
    \includegraphics[width=0.95\columnwidth]{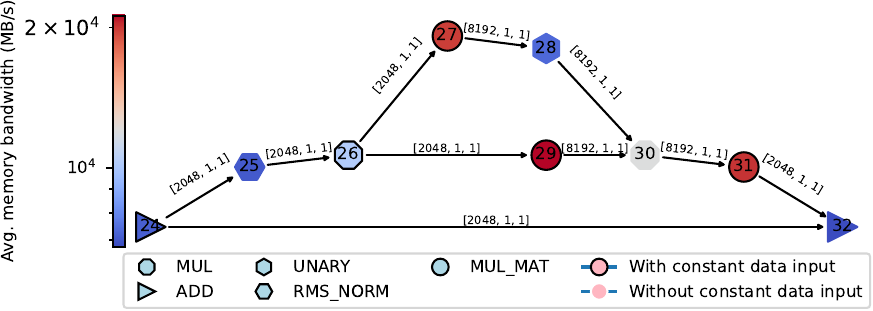}
    \vspace{1mm}
    \caption{Graphical view of an FFN for LLaMA-3.2-1B, in terms of memory bandwidth.}
    \label{fig:graph_view}
\end{figure}

The structural information is obtained by iteratively querying the source nodes' addresses of each node (lines 15-20). The encountered source nodes that were not previously added to the graph represent constant tensors---typically weight matrices---that are not involved in any operator computation. For each existing dependency between two nodes, we add a directed edge to the graph. A graph processing package can help generate and plot the graph, e.g., networkx~\cite{networkx}. Figure~\ref{fig:graph_view} depicts a ProfDAG of LLaMA3.2-1B-F16 running with two Cortex-A76 CPU cores on Orangepi 5 Ultra w.r.t. memory bandwidth.
{%
\footnotesize
\setlength{\textfloatsep}{5pt} 
\SetAlCapFnt{\footnotesize}
\SetAlCapNameFnt{\footnotesize}
\SetAlFnt{\footnotesize}
\begin{algorithm}[t]\label{alg:profdag}
    % \SetKwInput{KwPar}{Parameter}
    \caption{ProfDAG Retrieving.}
    \label{alg:profdag}
    \KwIn{
    $df_{raw}$: a table with all raw buffer entries\newline
    $n_{tar}$: target decoding iteration
    }
    \KwOut{$graph$: a ProfDAG given of LLM}
    % $df\_group_{tid} \gets \texttt{groupby}(\texttt{tid})$
    $dfs\_ops \gets \{ \}$;\\
    \ForEach{$df_{tid}$ in $df_{raw}.groupby(\texttt{tid})$}{
        $df_{tid} \gets df_{tid}.sorted(by=timestamp)$;\\
        % $df_{tid}^{iter} \gets df_{tid}[n_{iter} == n_{tar}]$;\\
        $df\_ops_{tid}^{iter} \gets df_{tid}[func == op \land n_{iter} == n_{tar}]$
        % $df\_ops_{tid}^{iter} \gets df_{tid}[func == op\_forward\_pass]$ 
        % \tcp*{filter out the operator functions}
        $dfs\_ops \gets dfs\_ops \cup \{df\_ops_{tid}^{iter}\}$;\\
    }
    $graph \gets DAG()$\\
    $df\_ops \gets dfs\_ops[0]$\tcp*{One thread's result}
    \For{$i \gets 1$ \KwTo $\texttt{len}(df\_ops)$ \textbf{step} 2\tcp*{At start}}{
    % \ForEach{$entry$ in $df\_ops$\tcp}{
        $op_{cur} \gets df\_ops['addr']$\\
        % $ts_{start} \gets \min\{df[i]['ts'] \mid df \in dfs\_ops\}$;\\
        % $ts_{end} \gets \max\{df[i+1]['ts'] \mid df \in dfs\_ops\}$;\\
        $op_{cur}['ts'] \gets get\_end\_to\_end\_time(dfs\_ops, i)$;\\
        % $node_{cur}['pmc'] \gets \sum\{df[i+1]['pmc'] - df[i]['pmc'] \mid df \in dfs\_ops\}$;\\
        $op_{cur}['pmc'] \gets sum\_pmc\_diff(dfs\_ops, i)$\\
        $op_{cur}['info'] \gets get\_op\_info(dfs\_ops, i)$\\
        % $node_{cur}[\{'type', 'name', 'dims'\}] \gets dfs\_ops[0][\{'type', 'name', 'dims'\}]$;\\
        $op_{cur}['order'] \gets i$;\\
        $graph.add\_node(op_{cur})$;\\
        \ForEach{$op_{pre}$ in $dfs\_ops[0]['src']$}{
            \If{$op_{pre}$ not in $graph.nodes$}{
                % $node_{pre}['type'] \gets $ 'constant';\\
                $graph.add\_node(op_{pre})$;\\
            }
            $graph.add\_edge(op_{pre}, op_{cur})$;
        }
    \KwRet{graph};
    }
\end{algorithm}
}%

% \bsi{If we want to distinguish the output DAG from the input workload DAG model, we can call the output something like "runtime characterization graph (RCG)". To have a bit of formalism, we need to introduce the definition of the node shapes, node colors, etc., used in this RCG model.}

% \begin{itemize}
%     \item Besides the model architecture, ProfDAG also enables to visually compare quantitative traits of different operators in a model or across models. (One with time, one with PMC)
%     \item An algorithm about finding the pair of uprobe and uretprobe, and how to retrieve the DAG architecture.
% \end{itemize}

\subsection{Timeline (ProfTime) view}
To visualize the temporal execution behavior over the entire workflow, {\sys} can generate a timeline view (ProfTime) for token-level, graph-level, and operator-level activities, as well as the scheduling behaviors across different threads. For instance, ProfTime can show how the computation graph is partitioned and scheduled over different backends.

To reveal the underlying scheduler behaviors, {\sys} captures all the thread IDs of the inference. From the context given by \texttt{sched\_switch} and \texttt{sched\_wakeup}, we identify the next status of the working thread~\cite{abaza2024trace}. If the working thread is switched out with \texttt{prev\_state} equal to 0, it is preempted and turns "idle". If the working thread is woken up or switched out with \texttt{prev\_state} equal to 1, the thread turns "runnable". Working thread that switches in is considered "running". Also, the processor ID is appended to each event's name.
By converting the raw trace data into the Chrome Trace Event Format~\cite{trace_event_format_chrome}, ProfTime can visualize the timeline with the help of Perfetto~\cite{perfetto_tracing}.

% Besides, it also records the timing of thread switches and wake-ups, from which we could identify when the threads are running 
% \begin{itemize}
%     \item introduce trace event format
%     \item a chrome-trace showing token-level, graph-level, and operator-level behavior.
%     \item a chrome-trace showing the thread 
%     \item graph-level
% \end{itemize}

\subsection{Statistical (ProfStat) view}
To analyze the detailed characteristics that inherently impact the inference performance and to uncover the correlation among them, we propose ProfStat view in three dimensions.
\emph{Across tokens:} By varying the number of tokens of the prompt and the generated context, we analyze the variation of the time of both prefill and each decoding iteration, as well as the influential operators. \emph{Per-operator type:} Given a fixed type of the operator, e.g., matrix multiplication (MatMul), the actual performance also depends on the dimensions of the input and output tensors, which are determined by the model and the inputs. PMCs' values on CPU and profiling results on the other backend can further help identify the performance bottleneck of a certain type of operator. \emph{Across experts:} For MoE models, ProfStat can help plot the activated experts over iterations and derive some insights about expert reuse.
% \begin{itemize}
%     \item Token
%     \item Operator type (e.g. for Mul Mat)
%     \item MoE, attribute for each expert: activated numbers, distance between the previous activation.
    
% \end{itemize}

% \newpage
\section{Evaluations}
\label{sec:eval}
%3 columns
\begin{table}[]
    \centering
    %\scriptsize
    \resizebox{\columnwidth}{!}{
        \begin{tabular}{l|l|l|l}
        
            \toprule
             Device &  SoC & RAM & OS\\
             \midrule
             Orange Pi 5 Plus & RK3588 & 32G LPDDR4/4X & OpenHarmony 5.1.0 \\
             % Orange Pi 5 Pro & RK3588 & 16G LPDDR5 & Ubuntu 22.04.5 \\
             Orange Pi 5 Ultra & RK3588 & 8G LPDDR5 & Ubuntu 22.04.5 \\
             RUBIK Pi 3 & QCS6490 & 8G LPDDR4X & Ubuntu 24.04 \\
             \bottomrule
        \end{tabular}
    }
    \caption{List of test devices and their configurations.}
    \label{tab:devices}
\end{table}
We evaluate {\sys} on three devices listed in Table~\ref{tab:devices}.
While {\sys} is developed using \texttt{libbpf} on the Orange Pi 5 Plus due to a light-weight design of OpenHarmony, \texttt{BCC} is used on other devices that have more features.

\comment{
\helper{
\begin{itemize}
    \item Talk about different setups on which it is tested. 
    % On Orangepi 5 pro, Ubuntu
\end{itemize}
}
}
\subsection{Overhead analysis}

\paragraph{Overhead of existing ML profilers.}
llama.cpp has a built-in function called \texttt{ggml\_graph\_dump\_dot} that writes the computational graph representing an LLM into a \emph{.dot} format file.
The output contains only the architectural information of a model without any performance data.
This functionality involves making one forward pass through the model, however, the time for that pass is \emph{13\%} longer than when it is not used, i.e., it incurs a \emph{13\% overhead in execution time}.
As a sophisticated machine learning library, ONNX Runtime~\cite{onnxruntime} also supports a profiling tool, whose time overhead is around 8\% of the model execution time through our preliminary experiments.
% As a sophisticated machine learning library, ONNX Runtime also supports a profiling tool, yet it can increase the execution time by 8\% on a normal neural network without reporting memory usage.\bs{citation?}
% Hence, the tracing or logging mechanism implemented inside llama.cpp is not useful for runtime profiling and collecting performance data. \bz{A bit inconsistency as we are using OpenCL Profiling}

\paragraph{Overhead of \sys.}
To analyze the online overhead caused by {\sys}, we used \emph{bpftool} to collect the cost of every probe. Besides, as the decoding stage is memory-intensive, the message passing and logging of the tracer can also lead to a degradation of the decoding speed. Table~\ref{tab:overhead} summarizes the relative decrease in the decoding speed and the average CPU load costed by the probes w.r.t. the runtime (e.g., we observed the CPU load when using 4 threads is always 400\%, then the average CPU load of the probes on each core is equal to $\sum(t_{probe}) / (t_{runtime} \times 4)$). Specifically, there is no impact on the prefill stage. 

As shown, {\sys} supports three major controlling flags that are configured in the initialization phase. \textit{Str.} refers to tracing the address, dimensions, and also source tensors of each operator. \textit{PMC} refers to reading the PMC value at each probe handler. \textit{P-buf.} means using the perf buffer to be aware of the missing events, while the ring buffer is chosen for lower memory overhead. The average CPU load of the probes is basically negligible. In BCC, the decoding speed degradation can vary from 2.8\% to 4\%. As a C-based library, libbpf has the lowest speed drop of 1.7\%, while some features are still not supported. Besides, the results are collected without switching off any type of operator. When tracing only at the token and graph level, the speed decrease is only 0.1\%.
%Also, llama.cpp supports a built-in function called \texttt{ggml\_graph\_dump\_dot}, plotting the computation graph into a ".dot" format file. Whereas, it only contains the structural information without any detailed metrics, and needs instrumentation inside the source code. By adopting it inside the warm-up stage, i.e., one empty decoding iteration before the prefill stage, and generating one token only, it adds about 13\% additional execution time, compared to the one without it.\bz{Please check if this overhead makes sense, or I can profile the time of this function.}
{
\footnotesize
\begin{table}[t]
    \centering
    \resizebox{0.82\columnwidth}{!}{
        \begin{tabular}{c|ccc|cc}
             Plat. & Str. & PMC & P-buf. & Speed dec. & CPU load P.\\
             \midrule
            \multirow{8}{*}{\rotatebox{90}{BCC}} 
            & \cmark & \cmark & \cmark & 4.0 & 0.70\\
            & \xmark & \cmark & \cmark & 3.6 & 0.66\\
            & \cmark & \xmark & \cmark & 3.4 & 0.69\\
            & \cmark & \cmark & \xmark & 3.1 & 0.62\\
             % & \xmark & \cmark & \xmark & \cmark & 4.4 & 0.75\\
             & \xmark & \cmark & \xmark & 3.3 & 0.63\\
             & \xmark & \xmark & \cmark & 3.3 & 0.63\\
             & \cmark & \xmark & \xmark & 3.0 & 0.62\\
             % & \xmark & \cmark & \xmark & \xmark & 4.5 & 0.72\\
             & \xmark & \xmark & \xmark & 2.8 & 0.61\\
             \midrule
             \multirow{2}{*}{\rotatebox{90}{libbpf}} 
             & \cmark & \xmark & \xmark & 2.2 & 0.41 \\
             & \xmark & \xmark & \xmark & 1.7 & 0.41 \\
             \bottomrule
             
             % & \xmark & \xmark & \xmark & \cmark & 4.9 & 0.81\\
             % & \xmark & \xmark & \xmark & \xmark & 4.5 & 0.73\\
        \end{tabular}
    }
    \caption{Overhead analysis with 2 and 4 Cortex-A76 cores on Orangepi 5 Pro (BCC) and Orangepi 5 Plus (libbpf). \textit{Str.}: parsing the operator structure. \textit{PMC}: enabling PMC reading. \textit{P-buf.}: using perf buffer. \textit{Speed dec.}: the relative decrease of the decoding speed in percentage. \textit{CPU load P.}: the average CPU load of the probe handlers in percentage.}
    % \bzi{The value might be wrong, because I might lose some experiments. Otherwise the last config should have the minimum overhead.}
    % \bsi{relative decrease means percentage? How to interpret the cpu load? (Maybe this will be clear after adding the descriptions in the text)}
    \label{tab:overhead}
\end{table}
}
% Experiment setting:
% \begin{itemize}
%     \item 3 models: Llama-3.2-1B, Deepseek-R1-Distill-Qwen-1.5B, Gemma2-2B
%     \item thread settings: 1 to 4 threads bound on Cortex-A76 cores
%     \item configs: 12 configurations, as when Str. is on, OPs should be always on.
% \end{itemize}
% \bz{Memory overhead?}
% \helper{
% \begin{itemize}
%     \item table showing configurability of our tracing framework as well as the overheads for different configurations
%     \item from OH on Opi5 plus, give the maximum speed degradation for the same model and inference setting. comment about the tracing configuration for the libbpf experiment. (dhiman) 
%     \item give some numbers on ggml\_graph\_dump\_dot (drawback: needs re-compilation and running the inference, high overhead)
%     \item onnx overheads (dhiman)
% \end{itemize}
% }

% \bs{What do the numbers mean in the table?}

% \bsi{Take a look at Fig. 4 and Fig. 5 in the eInfer paper and Fig. 2 in the eGPU paper to check how they present their overheads.}

\subsection{Profile-driven insights}

\subsubsection{ProfDAG visualization}
\begin{figure}
    \centering
    \includegraphics[width=0.95\columnwidth]{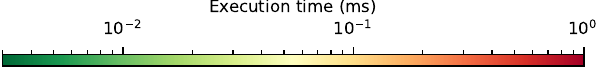}
    \subfloat[LLaMA3.2-1B]{
        \includegraphics[width=0.9\columnwidth]{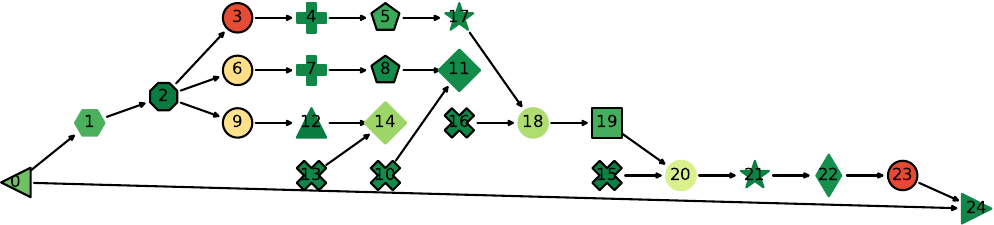}
    }
    \quad\subfloat[Qwen2.5-1.5B]{
        \includegraphics[width=0.9\columnwidth]{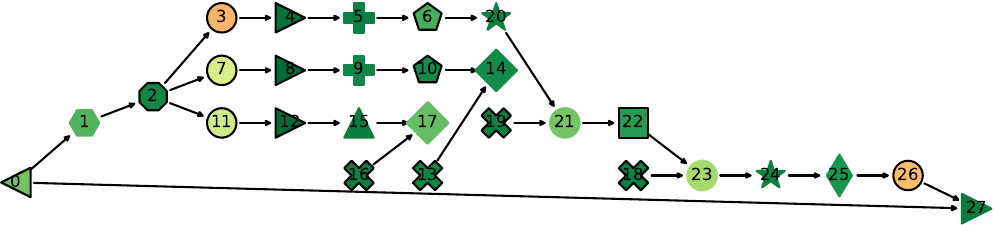}
    }
    \quad
    \subfloat[Gemma2-2B]{
        \includegraphics[width=0.9\columnwidth]{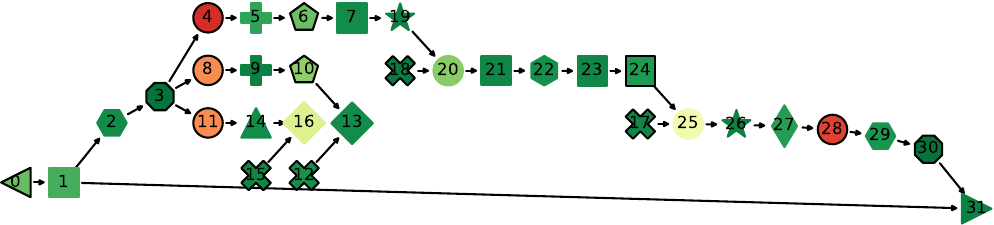}
    }
    \quad
    \includegraphics[width=0.95\columnwidth]{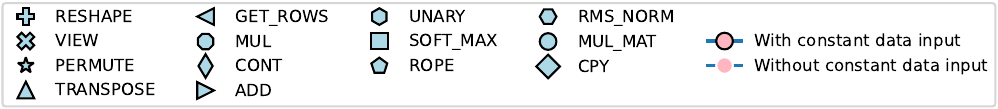}
    \caption{Architecture differences in the self-attention of different LLMs.}
    \label{fig:graph_arch}
\end{figure}
We apply ProfInfer to obtain ProfDAG of three LLMs, namely LLaMA3.2-1B, Qwen2.5-1.5B, and Gemma2-2B. Using them, we infer the following: 
(i)~The differences in the self-attention implementation of them can be observed in Figure~\ref{fig:graph_arch}. Compared to LLaMA3.2-1B, we see that inside one transformer layer Qwen2.5-1.5B has an ADD (e.g., 4,8, or 12 denoted by \RightTriangle) after three MUL\_MAT (i.e., 3, 7, and 11 denoted by \Circle). Further, Gemma2-2B has six more operators compared to LLaMA3.2-1B including three SOFT\_MAX (\Square) and one each of UNARY (\RHexagon), RMS\_NORM (\Hexagon), and MUL (\Octagon). 
(ii)~ProfDAG shows the execution order of tensor operators during an LLM inference. Here, we see that MUL\_MAT 3, 6, and 9 in LLaMA3.2-1B do not execute consecutively despite having the same intermediate tensor as input. A similar observation can be made for Qwen2.5-1.5B and Gemma.
(iii)~In all three ProfDAGs, we see that two MatMuls (e.g., MUL\_MAT 4 and 28 in Gemma2-2B) are the most heavy operations in self-attention. Also, the same operation takes longer in LLaMA3.2-1B (MUL\_MAT 3) and Gemma2-2B (MUL\_MAT 4) compared to Qwen2.5-1.5B (MUL\_MAT3). In Figure~\ref{fig:graph_arch}, the color of a node conveys the execution time, however, any other metric (e.g., Figure~\ref{fig:graph_view} shows average memory bandwidth utilization) can be visualized as well.

\subsubsection{ProfTime visualization}
\begin{figure}
    \centering
    \includegraphics[width=\columnwidth]{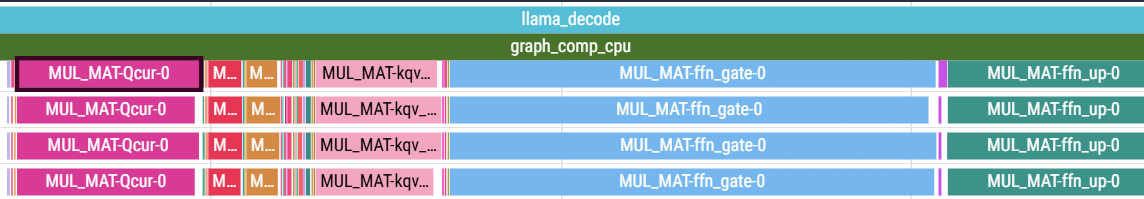}
    \caption{Timeline view in Perfetto. Only intra-operator parallelism is observed. (LLaMA3.2-1B-F16 with 4 threads.)}
    \label{fig:timeline_op_seq}
\end{figure}

We run an inference based on LLaMA3.2-1B-F16 over llama.cpp with 4 CPU threads on Orange Pi 5 Ultra.
{\sys} generates a ProfTime (timeline) visualization of it using which we can infer the following:
(i)~Figure~\ref{fig:timeline_op_seq} shows that llama.cpp runs tensor operations one after the other. However, one tensor operation, e.g., MatMul, can be run by multiple threads. That is, it exploits intra-operator parallelism but there is no inter-operator parallelization. 
(ii)~It can be further observed in the extracted segment, that MatMuls dominate other operations which holds for the entire inference as well, i.e., more than 97\% of TTFT and TPOT is spent on MatMuls. 

\begin{figure}
    \centering
    \subfloat[Activation function]{
        \includegraphics[width=\columnwidth]{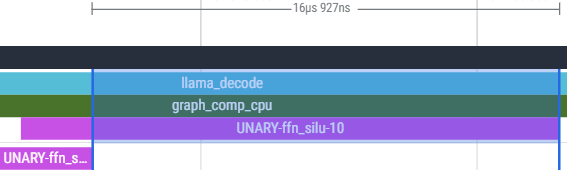}
    }
    \quad
    % \vspace{-1mm}
    \subfloat[Matrix multiplication]{
        \includegraphics[width=\columnwidth]{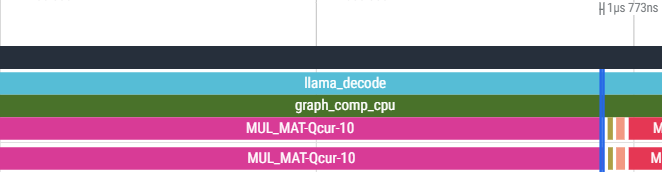}
    }
    % \vspace{-1mm}
    \caption{(Un)balanced operator load across threads.}
    \label{fig:timeline_op_async}
\end{figure}
Figure~\ref{fig:timeline_op_async} demonstrates that the computational load is not uniformly distributed among inference threads for all operators. Here, LLaMA3.2-1B-F16 is run by two threads.
In (a), we see that most computations in the activation function are performed by one thread while the other is idle for almost 80\% of the operator's runtime. 
However, (b) shows that the computations in a MatMul can be distributed almost equally between two threads.

\begin{figure}
    \centering
    \subfloat[Rockchip NPU backend (graph-level).]{\includegraphics[width=\columnwidth]{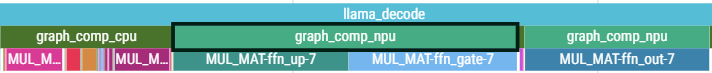}}
    \quad
    \subfloat[CLBlast backend for ARM Mali GPU (operator-level).]{\includegraphics[width=\columnwidth]{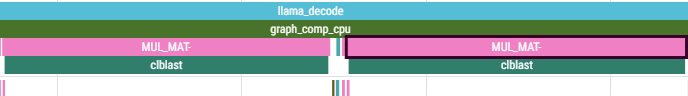}}
    \quad
    \subfloat[OpenCL backend for Adreno GPU (graph-level and kernel-level observability combined with built-in OpenCL profiling).]{\includegraphics[width=\columnwidth]{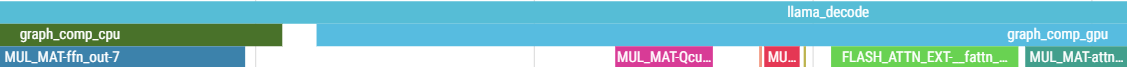}}
    \caption{Timeline view of graph and operator execution on different backends.}
    \label{fig:multi_backend}
\end{figure}
Figure~\ref{fig:multi_backend} demonstrates ProfTime's capability to show how LLM inferences run across multiple backends.
(a) and (c) shows graph partitioning in llama.cpp where an inference is decomposed into multiple computation graphs that run on different backends, i.e., CPU and Rockchip NPU backends in (a) and CPU and OpenCL (for Adreno GPU) backends in (c). 
Here, the backend is determined at the graph level.
(b)~shows nested backends, i.e., the CLBlast backend is implemented under the CPU backend. CLBlast backend is at the operator level, e.g., we see that when a MatMul is encountered by the CPU backend, it offloads the operation via CLBlast to ARM Mali GPU on Orange Pi 5 Ultra.
In all three cases, ProfTime shows the exact backend at the graph level.
(a) and (b) further shows the operator's runtime directly from our tracing results, while for OpenCL, we need to turn on the flag \texttt{GGML\_OPENCL\_PROFILING} to enable the built-in profiling tool and combine it with our results.
% which is not clear in (c) when an operator is executed by the OpenCL backend.
The reason is that the OpenCL backend selects the OpenCL kernels for each operator and enqueues them into a command queue (CQ), from which the kernels are dispatched asynchronously. 
% To enable observability into OpenCL kernel execution, we need to incorporate OpenCL profiling in our framework, which is future work.
% In (c), we see that the CPU backend performs the Mat Mul in the Language Modeling (LM) head, as the accumulative kernel size exceeds the maximum allocation limit allowed by OpenCL on the Adreno GPU.
% whose large dimensionality is not supported by OpenCL.

%The backend is determined at the graph level. Here, ProfTime shows the exact backend that runs an operator besides the operator's runtime.
%(b)~shows nested backends, i.e., the CLBlast backend is implemented under the CPU backend. CLBlast backend is at the operator level, e.g., we see that when a matrix multiplication is encountered by the CPU backend, it offloads the operation via CLBlast to ARM Mali GPU on Orange Pi 5 Ultra.
%(c)~shows an OpenCL backend that selects an OpenCL kernel for each supported operation and enqueues it into the OpenCL command queue (CQ). The kernels are dispatched from the CQ asynchronously. To enable observability into OpenCL kernel execution, we need to incorporate OpenCL profiling in our framework, which is future work. Besides the OpenCL backend, we see that the CPU backend performs the matrix multiplication in the Language Modeling (LM) head, whose large dimensionality is not supported by OpenCL.

\begin{figure}
    \centering
    \includegraphics[width=\linewidth]{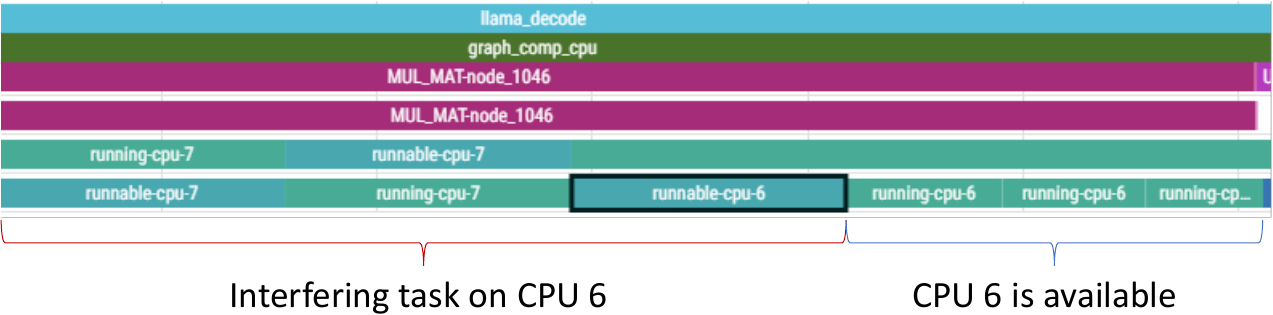}
    \caption{Timeline view of the thread and operator tracing of running LLaMA3.2-1B on 2 Cortex-A76 cores with a interfering task on one core.}
    \label{fig:interfere}
\end{figure}
ProfTime can pinpoint if an operator is delayed due to interference from a high-priority task, as shown in Figure~\ref{fig:interfere}. 
Here, two inference threads are bound to CPUs 6 and 7, while the interfering thread is running on CPU 6.
We see that the two threads share CPU7 while CPU6 is occupied. After CPU6 becomes available, both threads run simultaneously on the two CPUs.
The experiment shows how ProfTime can help debug unexpected delays in the inference.

\comment{
\helper{
(i)~show sequential processing of tensor operators; (ii)~show how the graph computing across multi-backend works; (iii)~show multi-threaded execution of an operator, e.g., mul\_mat; (iv)~show single-threaded execution of an operator, e.g., rope; (v)~trace sched\_switch and show interference. 
}
}

\subsubsection{ProfStat visualization}
\begin{figure}
    \centering
    \includegraphics[width=\columnwidth]{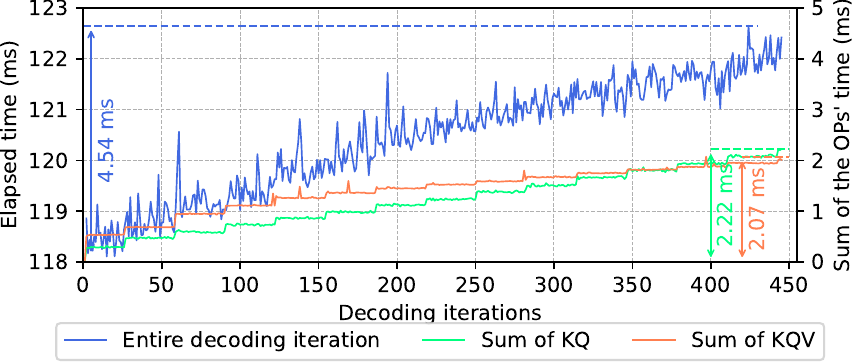}
    \caption{The elapsed time for generating one token and the sum of the \textit{KQ} and \textit{KQV} across layers in each decoding iteration. The model is LLaMA3.2-1B, which runs with 2 threads. It outputs 450 tokens. The increase in the iteration time is caused by \textit{KQ} and \textit{KQV}.}
    \label{fig:kv_cache}
\end{figure}
To analyze the performance across the tokens, we look into the profiling results of the CPU backend, as it still provides the near-optimal performance.
In the decoding stage, the KV-cache mechanism reduces some computation, resulting in stable decoding throughput. However, as the KV cache grows, the elapsed time for each iteration still increases, as shown in Figure~\ref{fig:kv_cache}. After analyzing all the operators, we find that the operators \textit{KQ} and \textit{KQV} contribute the most to the degradation, with a step-wise time increase as the context length grows.

\begin{figure}
    \centering
    \includegraphics[width=0.9\linewidth]{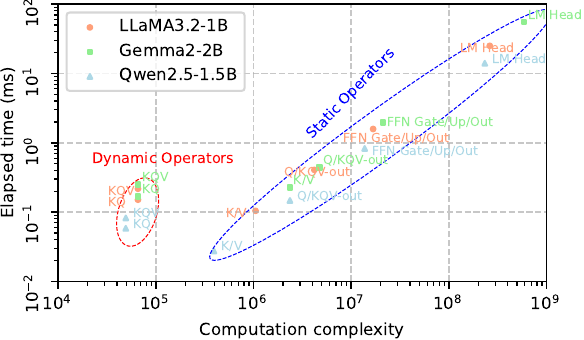}
    \caption{The execution time of the operator matrix multiplication in the first decoding iteration with 2 Cortex-A76 CPUs as the backend. The x-axis represents the computational complexity, i.e., $\mathcal{O}(M \times N \times K \times H)$, given a matrix multiplication $A_{(M,K)} \times B_{(K,N)}$. $KQ$ and $KQV$ also introduce the dimension of attention heads $H$. $M$ is 1 in the decoding stage.}
    \label{fig:op_mat_mul}
\end{figure}

Looking further into all MatMul operators in a single decoding iteration, all of them are matrix-vector multiplications, some of which with a broadcasting dimension of attention heads (\textit{KQ} and \textit{KQV}). Except for these two operators, the runtimes of the others are proportional to the computation complexity, as shown in Figure~\ref{fig:op_mat_mul}. Qwen2.5-1.5B has a slightly lower intercept and thus a faster speed for an operator with the same dimension, which is possibly caused by the smaller hidden dimension (1536).

\begin{figure}
    \centering
    \includegraphics[width=0.95\linewidth]{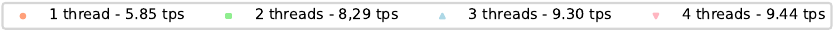}
    \subfloat[Memory access]{
        \includegraphics[width=0.47\columnwidth]{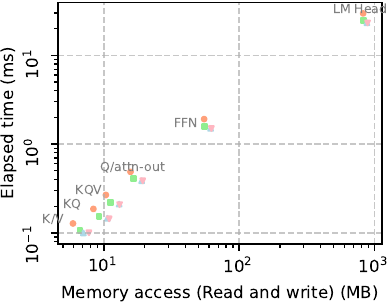}
        \label{fig:mem_access}
    }
    \subfloat[Backend bound]{
        \includegraphics[width=0.47\columnwidth]{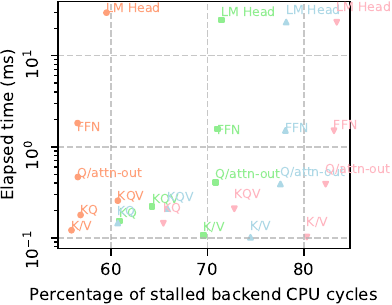}
        \label{fig:cpu_cycles}
    }
    \caption{Memory access and stalled cycle ratio of matrix multiplications in the decoding stage with different numbers of thread settings.}
    \label{fig:op-pmc}
\end{figure}
As decoding is memory-intensive, we read two per-core counters of Cortex-A76 CPUs to calculate the data transferred between the last-level cache and main memory. Figure~\ref{fig:mem_access} illustrates the elapsed time of each MatMul w.r.t. the memory access. We see that increasing the number of threads can improve the speed dramatically, only up to 2 threads. More threads can solely speed it up slightly, yet reach the memory bandwidth bottleneck and could cause some unnecessary cache misses. Figure~\ref{fig:cpu_cycles} also shows the ratio of stalled backend cycles w.r.t. the total CPU cycles. It implies that 4 threads can cause more than 80\% of stalled cycles, which is a waste of resources. In conclusion, due to the linear correlation of the MatMuls, the decoding speed of a dense model can be predictable given the hyperparameters.

In the prefill stage, many existing techniques can leverage the parallelism well with heterogeneous backends~\cite{song2024powerinfer, xu-2025}, while CPU can also utilize some libraries to better exploit the cache locality for MatMul, such as BLIS~\cite{BLIS1}. By implementing BLIS on 4 Cortex-A76 cores, the prefill speed is twice as fast as when it is not used. By reading the PMCs l3d\_cache\_refill and mem\_access\_wr, we find that it reduces the original memory access by 75\%.

\begin{figure}
    \centering
    \includegraphics[width=0.9\columnwidth]{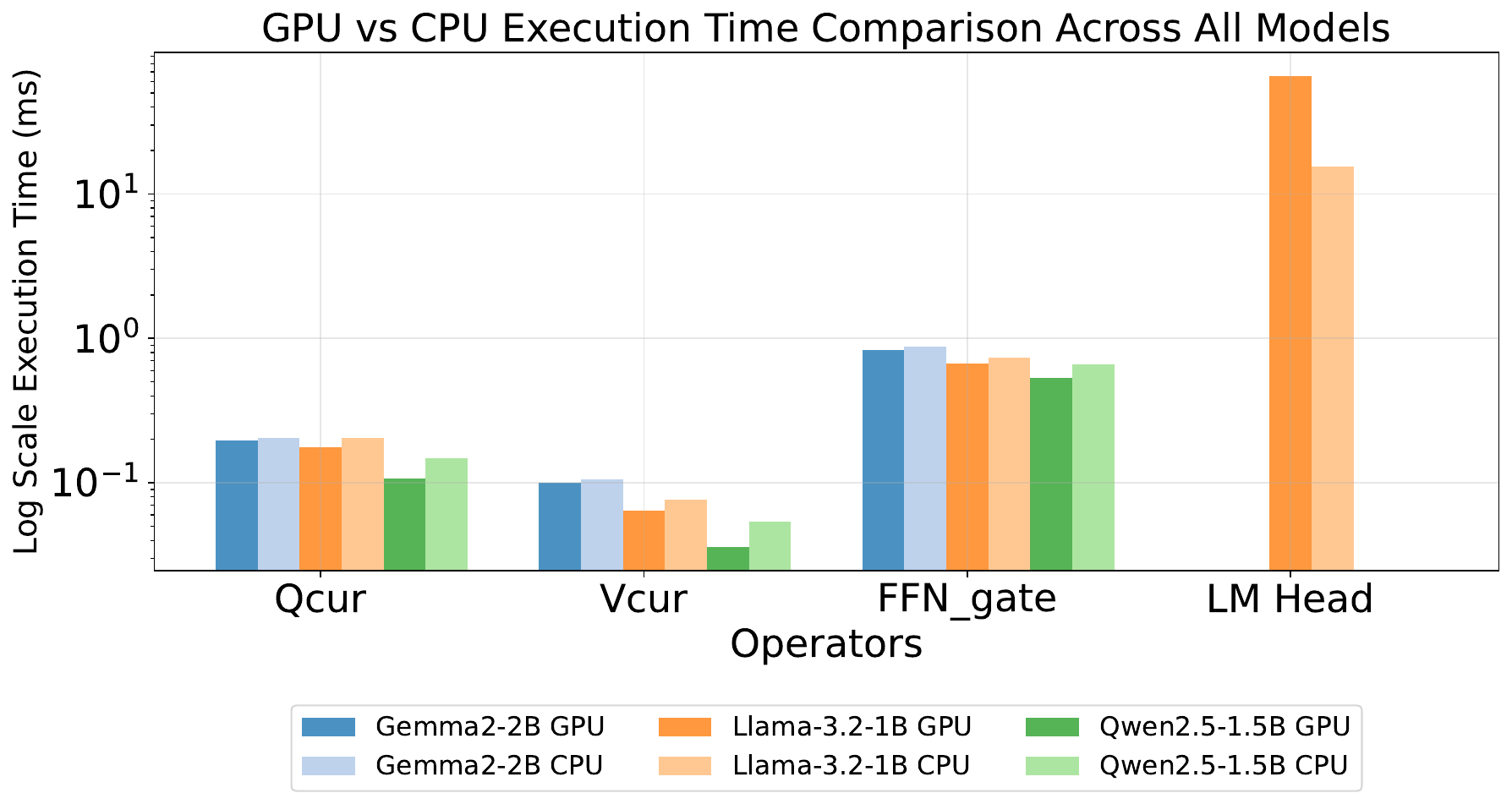}
    \vspace{1mm}
    \caption{The comparison of matrix multiplication operators of the GPU and CPU on Rubik Pi. Both settings are at their peak decoding speeds.}
    \label{fig:gpu_cpu_comp}
\end{figure}
When offloading the layers to the accelerators, {\sys} can also provide a fine-grained comparison on different backends. Taking the Adreno GPU of the Rubik Pi as an example, Figure~\ref{fig:gpu_cpu_comp} depicts the execution time of all the static MatMuls on both the CPU (with peak performance at 4 threads) and the GPU with these 3 LLMs of 4-bit quantization. For Gemma2-2B and Qwen2.5-1.5B, \textit{LM Head} cannot be offloaded due to the size limit of OpenCL kernels. We observe that the GPU outperforms the CPU only in a certain range of dimensions. For the LM Head, whose matrix size is much larger, there is a dramatic drop in GPU performance. Thus, regardless of the data transfer between CPU and GPU, selective operator offloading based on the tensor dimension can potentially accelerate the inference speed.

\begin{figure}
    \centering
    \includegraphics[width=\columnwidth]{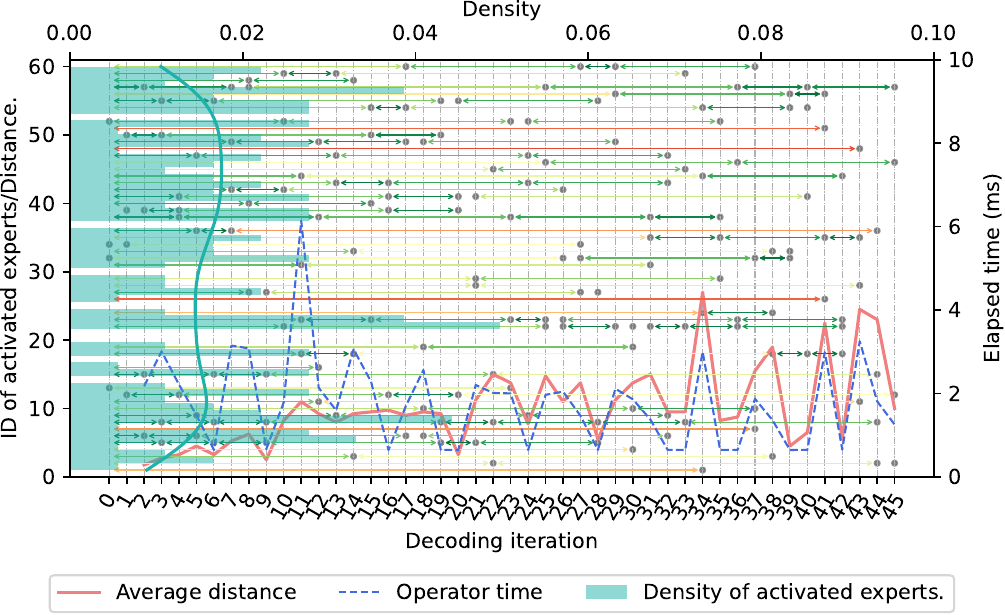}
    \caption{The activated experts of the \textit{ffn\_moe\_up-0} operator of Qwen1.5-MoE-A2.7B-Q4 (60 experts in total with 4 activated experts each time). Lower x-axis: each decoding iteration. Upper x-axis: density distribution of the activated experts. Left y-axis: ID of the experts. Right y-axis: Elapsed time of the operator. The average distance refers to the mean value of the distance between the current iteration and when the 4 experts were activated last time.}
    \label{fig:moe_one_op}
\end{figure}
Figure~\ref{fig:moe_one_op} shows a profiling result of a MatMul in the FFN layer of an MoE model, Qwen1.5-MoE-A2.7B~\cite{bai2023qwen}, which has 60 experts at each FFN layer and activates 4 of them each time. Even with 4-bit quantization, it still requires 8.9 GB of storage, which cannot be fully loaded into the main memory of the Orangepi 5 Ultra and is only runnable by using \texttt{mmap}. The lower x-axis refers to each decoding iteration, while the left y-axis refers to each expert ID. The top x-axis shows the density distribution of the activated experts over iterations, where we plot the frequency with which each expert is activated across the runs. The right y-axis represents the elapsed time. Inside the figure, we annotate the distance between two consecutive activated experts and calculate the average distance of the four experts of the current iteration. After a few iterations, we observe that the operator time is approximately proportional to the average distance. It indicates that the long-distance experts are likely to be evicted from the memory. By monitoring the major page faults and memory access, we conclude that the bottleneck of the MoE model inference is located in disk I/O rather than memory bandwidth.

\comment{
\helper{
Directions:
\begin{itemize}
    % \item \sout{LLaMA3.2-1B, Qwen2-1.5B architecture difference.}
    % \item Complexity, number of parameters, how the fine-grained tracing reflects how the architecture influence the decoding speed.
    % \item KV cache - for one operator and one token generation.
    \item Operator-level: linear-correlation, LM head take the maximum
\end{itemize}
Content
\begin{itemize}
    \item \sout{Talk about insights from graph visualization}
    \item \sout{Talk about insights from timing visualization (e.g., chrome visualization), how inference engine executes the graph, i.e., sequential; not all threads are used for all operations}
    \item Talk about our observations (performance counters compare with memory benchmarks) when we try to parallelize tensor operations
    \item Show some correlation between performance counters and operator timings
    \item Comment on the impact of KV cache and operator timings and TPT
    \item Insights about MoE execution
    %\item Insights about speculative decoding execution
    \item Insights about batched inference and continuous batching; which batch size works well and why?
    \item Talk about operator offloading based on profiling   
    \item run inference with different number of threads and observe how performance counter values change, can that be used to deduce memory boundedness
    \item run inference with different quantization and derive insights
    \item run inference with other task and draw insights from profiles    
\end{itemize}
}
}

% Overhead: the difference of TPS/TTFT
% Accuracy: the difference between timestamp in eBPF and timestamps of the function invocation
% The ts when the buffer is obtained in userspace between the ts of invocation

% \subsection{Discussion}
% \begin{itemize}
%     \item Memory benchmark
%     \item Graph partitioning
%     \item Online scheduling (potentially it could dynamically decide the execution order at runtime, by changing the tensor address obtained by the first run. But how could it help?)
% \end{itemize}
% \bs{I would prefer to do the discussion in a separate section, as this is not only the discussion about the evaluation/experiments, but also for the proposed design.}
% \input{sections/discuss}
% \newpage
\section{Related Works}
\label{sec:related}
%0.75 cols

% \bsi{I drafted a preliminary version of the related works, please check}

\paragraph{Built-in profilers in general-purpose ML frameworks.}
Profiling and observability are critical for identifying performance bottlenecks in machine learning inference engines.
Traditional frameworks such as TensorFlow Lite~\cite{tflite_profiler}, ONNX Runtime~\cite{onnxruntime_prof}, and TensorRT~\cite{tensorrt_profiler} offer built-in profilers to report operator execution time, memory allocation, and threading behavior. 
% LLMs exhibit distinct runtime characteristics, such as autoregressive decoding and dynamic KV-cache reuse, which are not fully captured by these profilers.
Recent LLM inference engines, such as vLLM~\cite{kwon2023efficient} and TensorRT-LLM~\cite{tensorrtllm} optimize batching and memory scheduling for high-throughput generation.
However, these profilers are typically intrusive, \textit{i.e.}, requiring runtime flags or compilation options. Moreover, their observability features remain coarse-grained, typically exposing aggregate throughput, token rate, and latency.
% without the capability to reveal fine-grained performance limitations, especially on mobile hardware with limited resources. 

\paragraph{eBPF-Based profiling and system tracing.}
Recent advances leverage eBPF for performance analysis in deep learning systems.
\emph{eInfer}~\cite{chu2025einfer} demonstrates fine-grained tracing for distributed LLM inference using eBPF to collect runtime call graphs without source modification.
Similarly, \emph{eGPU}~\cite{yang2025egpu} extends eBPF programmability to GPUs, and Craun et al.~\cite{craun2024eliminating} propose techniques to eliminate eBPF tracing overhead on untraced processes.
These studies highlight the promise of eBPF for low-intrusion telemetry, yet they stop short of providing semantic alignment between ML operators and hardware performance counters.
Moreover, existing tools are evaluated primarily on server-class CPUs or GPUs and do not consider the constraints of on-device LLM inference and dynamic execution behaviors (\textit{e.g.}, mixture-of-experts routing or speculative decoding).

\paragraph{Mobile and on-device inference profiling.}
Mobile inference frameworks prioritize efficiency under tight resource budgets.
For example, MNN~\cite{jiang2020mnn} provides a lightweight cross-platform runtime for DNNs on smartphones and embedded systems, focusing on latency and energy metrics rather than detailed operator introspection.
Recent benchmarks such as PalmBench~\cite{lipalmbench} evaluate compressed and quantized LLMs across mobile devices, emphasizing throughput and accuracy.
However, these evaluations still lack \emph{fine-grained}, \emph{non-intrusive} profiling, which prevents developers from understanding how different operators and dynamic execution behaviors impact on-device LLM performance.

In summary, prior profilers either (i) require intrusive instrumentation within ML runtimes, (ii) provide only coarse metrics through serving frameworks, or (iii) apply eBPF tracing without integrating operator semantics or supporting mobile hardware.
\sys bridges these gaps by combining eBPF-based dynamic probes with operator-level performance-counter monitoring.
It achieves non-intrusive, fine-grained observability for LLM inference, correlating runtime events with hardware behavior and enabling detailed visualization of model execution on mobile platforms.

% \emph{eInfer}~\cite{chu2025einfer} mainly focuses on syscalls and driver-level activities and cannot perceive information at the model level, given the fact that operators of the same type can exhibit distinct performance depending on their location and dimension within the model.

\section{Conclusion and Discussion}
\label{sec:conclusion}
We propose a light-weight LLM inference profiling tool based on eBPF for mobile systems. By extracting the core functions of the software stack of llama.cpp, combined with hardware observabilities through performance monitoring counters, we derive fine-grained profiling results with three types of views to identify the performance bottlenecks.

As future work, based on the operator-level profiling results, we could implement graph partitioning to further exploit the inter-tensor parallelism for the decoding stage across heterogeneous backends. Additionally, we could utilize a memory benchmark~\cite{zuepke2024mempol, pradhan2025predictable} that reveals the maximum memory access capability to better understand the performance limit of the hardware. 
% \textbf{Memory-benchmark assistant bottleneck identification.} 

%0.25 cols

% this must go after the closing bracket ] following \twocolumn[ ...

% This command actually creates the footnote in the first column
% listing the affiliations and the copyright notice.
% The command takes one argument, which is text to display at the start of the footnote.
% The \mlsysEqualContribution command is standard text for equal contribution.
% Remove it (just {}) if you do not need this facility.

% \newpage

\bibliography{main}
\bibliographystyle{mlsys2025}

%%%%%%%%%%%%%%%%%%%%%%%%%%%%%%%%%%%%%%%%%%%%%%%%%%%%%%%%%%%%%%%%%%%%%%%%%%%%%%%
%%%%%%%%%%%%%%%%%%%%%%%%%%%%%%%%%%%%%%%%%%%%%%%%%%%%%%%%%%%%%%%%%%%%%%%%%%%%%%%
% SUPPLEMENTAL CONTENT AS APPENDIX AFTER REFERENCES
%%%%%%%%%%%%%%%%%%%%%%%%%%%%%%%%%%%%%%%%%%%%%%%%%%%%%%%%%%%%%%%%%%%%%%%%%%%%%%%
%%%%%%%%%%%%%%%%%%%%%%%%%%%%%%%%%%%%%%%%%%%%%%%%%%%%%%%%%%%%%%%%%%%%%%%%%%%%%%%
\appendix
% \section{Appendix}
% %
% Put anything that you might normally include after the references as an appendix here, {\it not in a separate supplementary file}. Upload your final camera-ready as a single pdf, including all appendices.

%%%%%%%%%%%%%%%%%%%%%%%%%%%%%%%%%%%%%%%%%%%%%%%%%%%%%%%%%%%%%%%%%%%%%%%%%%%%%%%
%%%%%%%%%%%%%%%%%%%%%%%%%%%%%%%%%%%%%%%%%%%%%%%%%%%%%%%%%%%%%%%%%%%%%%%%%%%%%%%

\end{document}